\documentclass[conference]{IEEEtran}
\IEEEoverridecommandlockouts
\usepackage{url}
\usepackage{cite}
\usepackage{amsmath, amssymb, amsfonts}
\usepackage{algorithmic}
\usepackage{graphicx}
\usepackage{textcomp}
\usepackage{enumerate}
\usepackage{xcolor}
\usepackage{subfigure}
\usepackage[T1]{fontenc}
\usepackage{multirow}
\usepackage[linesnumbered,ruled,vlined]{algorithm2e}
\usepackage{booktabs}    %导言区	
\usepackage{makecell}

\begin{document}

\title{MARE: Multi-Agents Collaboration Framework for Requirements Engineering}
 \author{
 \IEEEauthorblockN{Dongming Jin$^{1,2}$, Zhi Jin$^{1,2}$, Xiaohong Chen$^{3}$, Chunhui Wang$^{4}$}
    \IEEEauthorblockA{$^1$ School of Computer Science, Peking University, China}
    \IEEEauthorblockA{$^2$ Key Lab of High-Confidence of Software Technologies (PKU), Ministry of Education, China}
    \IEEEauthorblockA{$^3$ Shanghai Key Laboratory of Trustworthy Computing, East China Normal University, China}
    \IEEEauthorblockA{$^4$ College of Computer Science and Technology, Inner Mongolia Normal University, China}
    \IEEEauthorblockA{\texttt{dmjin@stu.pku.edu.cn}, \texttt{zhijin@pku.edu.cn}}
}

\maketitle

% 12点 完成方法 4页
% 3点 完成study design 5页
% 6点 完成result analysis 6页
% 9点 完成related work 7页
% 10点 完成conlucion discussion 7页半
%context, objectives, methods, and results and conclusions, and should not exceed 200 words.

\begin{abstract}
Requirements Engineering (RE) is a critical phase in the software development process that generates requirements specifications from stakeholders' needs. Recently, deep learning techniques have been successful in several RE tasks. 
However, obtaining high-quality requirements specifications requires collaboration across multiple tasks and roles. In this paper, we propose an innovative framework called \textit{MARE}, which leverages collaboration among large language models (LLMs) throughout the entire RE process. \textit{MARE} divides the RE process into four tasks: elicitation, modeling, verification, and specification. 
Each task is conducted by engaging one or two specific agents and each agent can conduct several actions. 
\textit{MARE} has five agents and nine actions. 
To facilitate collaboration between agents, \textit{MARE} has designed a workspace for agents to upload their generated intermediate requirements artifacts and obtain the information they need.
% share information, and these agents can upload their own generated intermediate requirements artifacts here and obtain the information they need from here.
We conduct experiments on five public cases, one dataset, and four new cases created by this work. 
We compared \textit{MARE} with three baselines using three widely used metrics for the generated requirements models. 
Experimental results show that \textit{MARE} can generate more correct requirements models and outperform the state-of-the-art approaches 
% on the three requirement models 
by 15.4\%. %, 23.9\% %, and 0.6\% %. 
For the generated requirements specifications, we conduct a human evaluation in three aspects and provide insights about the quality.

\end{abstract}


\section{Introduction}
Requirements Engineering (RE) is one of the most critical phases in the software development process. 
It aims to generate a well-defined set of requirements by eliciting, analyzing, specifying, and verifying the needs and expectations of stakeholders for a software system\cite{advancingrequirments}. 
As the complexity and scale of the software system\cite{frepa} continue to grow, developers will spend a lot of effort on various tasks in RE to create high-quality requirements models and write requirements specifications.
%developers costs lots of efforts to finish various tasks in requirements engineering, such as talking to get stakeholders' needs, analyzing requirement descriptions to create requirement model and specifying to write software requirement specifications (SRS). 

Nowadays, deep learning (DL) techniques have been successfully applied to automate various RE tasks\cite{dl4re}. 
For example, DL has been used to mine stakeholders' needs\cite{ShiCWB21}\cite{ShiXLWL020}, extract requirements model\cite{JinWJ23}\cite{WangHC22}, and deal with ambiguity in requirements\cite{EzziniA0SB21}. 
However, to obtain high-quality requirements specifications, different RE tasks need to be interwoven and iterated. The automation of only a few of these tasks limits further increase in effectiveness and efficiency.
%The previous DL-based models handle RE tasks separately and does not consider the collaboration between different tasks and roles, which limit the further improvement in effectiveness and efficiency.

Recently, large language models (LLMs) have achieved significant performance in the fields of software engineering (SE)\cite{llm4se} and natural language processing (NLP)\cite{llm4nlp}. 
There have also been some early explorations of the application of LLMs to RE\cite{jules}\cite{RuanCJ23}. 
For example, \cite{GorerA23} used LLMs to generate requirements elicitation interview scripts. \cite{FantechiGPS23} used ChatGPT to detect inconsistency in natural language requirements. 
\cite{RodriguezDC23} provided prompt strategies for requirements traceability. 
These efforts demonstrate the effectiveness of using LLMs for specific requirements tasks and do not take full advantage of LLMs for collaboration among RE tasks.
%These works deal directly with a specific requirement task using LLMs, and do not take full advantage of the collaboration between LLMs. 

Inspired by the multi-agents collaboration\cite{katzenbach2015wisdom} and team design, researchers started to investigate the possibility of designing the communicative agents mechanisms\cite{agentverse} for facilitating the accomplishment of complex tasks.
\cite{chatdev} presented a virtual chat-powered software development company to unveil fresh possibilities for integrating LLMs into the realm of software development. 
\cite{metagpt} proposed an meta-programming framework incorporating efficient human workflows into LLM-based multi-agent collaborations, allowing agents with human-like domain expertise to verify intermediate results and reduce errors. 
%For example, Hong et al. introduced a meta-programming framework for multi-agent collaboration based on LLMs. 
%These works are focused much on code generation and requirement elicitation, but they do not consider the modeling and verification for the generated requirements by LLMs, which can lead to missing or inconsistent requirements.
Studies show that multi-agent collaboration mechanisms are interrelated with specific tasks.
%For example, these works only foucs on requ

Along this line, in this paper, we propose a novel framework, named \textit{MARE} (\textbf{M}ulti-\textbf{A}gent collaboration for \textbf{R}equirements \textbf{E}ngineering) to automate the entire RE process based on collaboration between multiple tasks and roles. 
Specifically, given a very rough idea of requirements, \textit{MARE} autonomously and iteratively performs the basic tasks of RE, i.e., elicitation, modeling, verification, and specification, to achieve end-to-end requirements specifications generation. 
%To be specific, given a very rough idea about the software requirements, \textit{MARE} iteratively conducts the directly generates requirement model and requirement specification. It divides the RE process into four tasks: elicitation, modeling, verification and specification. 
%\textit{MARE} involves one or more LLM-based agents, such as stakeholders, collector, modeler, checker and documented. 
Each task is accomplished with one or two specific agents and \textit{MARE} contains five agents, i.e., the stakeholders, the collector, the modeler, the checker, and the documenter. 
Each agent is responsible for a requirements task and performs predefined actions to accomplish that task.
%Every agent is responsible for a certain requirement task and conducts the actions are the ability of agents. 
Nine actions are designed for this purpose.
\textit{MARE} also offers a shared workspace to these agents for supporting the collaboration among agents. 
%For example, the stakeholders agent is responsible for providing user requirements and it has \textit{SpeakUserStories} action and \textit{AnswerQuestion} action. The share work space that can store requirement artifacts (e.g. user stories, etc.) generated by every agent to facilitate effective communication and collaboration between agents.

We conduct extensive experiments to evaluate \textit{MARE}. 
(1) To evaluate its modeling ability, we compare \textit{MARE} with three baselines for automated requirements modeling on five public cases, one public dataset, and four new cases created in this paper. 
The three baselines are state-of-the-art (SOTA) methods for use case diagrams, problem diagrams, and goal models, respectively. 
Three widely used evaluation metrics (Precision(P), Recall(R), and F1) are used for the evaluation. 
Results demonstrate the impressive performance of \textit{MARE} on requirements modeling. 
In terms of F1, \textit{MARE(gpt-3.5-turbo)} outperforms the three SOTA baselines by up to 15.4\%, 23.9\%, and 0.6\%, respectively. 
(2) To evaluate its specification generation ability, we conduct a human evaluation to evaluate the quality of generated requirements specifications in three aspects (correctness, completeness, and consistency) and provide insights about their quality. 
(3) We conduct an ablation study comparing individual LLMs with \textit{MARE} on requirements modeling task. 
Results prove the contributions of multi-agents collaboration framework.

We summarize our contributions as follows.
\begin{itemize}
    % \item We formulate the problem of end-to-end requirements specification generation from a simple name of software system. 
    \item To the best of our knowledge, this is the first study exploring end-to-end automation of the entire RE process. This study unifies main tasks through LLMs, eliminating the need for specialized models at each task.
    \item We propose \textit{MARE}, a multi-agents collaboration framework for RE. By merely specifying a rough idea about the requirements, \textit{MARE} iteratively handles requirements elicitation, modeling, verification, and specification. 
    \item We conduct extensive experiments on nine cases and one dataset. Compared with three SOTA baselines, \textit{MARE} shows superiority in modeling. We further manually evaluate the generated specifications and provide insights about the quality of them in three aspects. 
\end{itemize}

In the remainder of the paper, Section II presents the approach. Section III sets up the experiments. Section IV describes the results and analysis. Section V introduces the related work. Section VI concludes our work.

  %1页半
% \input{chapters/2ProblemDefine} % 2页
\section{Approach}
This section presents the multi-agent collaboration framework, named \textit{MARE}, for end-to-end RE based on LLMs. 
%We formally define the overview of our MARE and describe its core components in the following, which involves 11 actions, 5 agents and 1 shared workspace.

\subsection{Overview} 
Requirements Engineering is characterized by discussions and brainstorming to define the scope, functionality, and quality of a product. 
The RE process is challenging because it usually consists of multiple tasks, e.g., requirements elicitation, requirements modeling, requirements verification, and specification. 
Traditionally, requirements engineers as well as stakeholders actively iteratively collaborate with each other to establish a common understanding of the envisioned product.
We investigate the possibility of automating the end-to-end RE process from a novel perspective by proposing a generative framework, i.e. to create a generative model $G(S|X)$ that generates a requirement specification $S$ based on a rough idea of requirements $X$. 

\begin{figure*}
    \centering
    \includegraphics[width=0.9\textwidth]{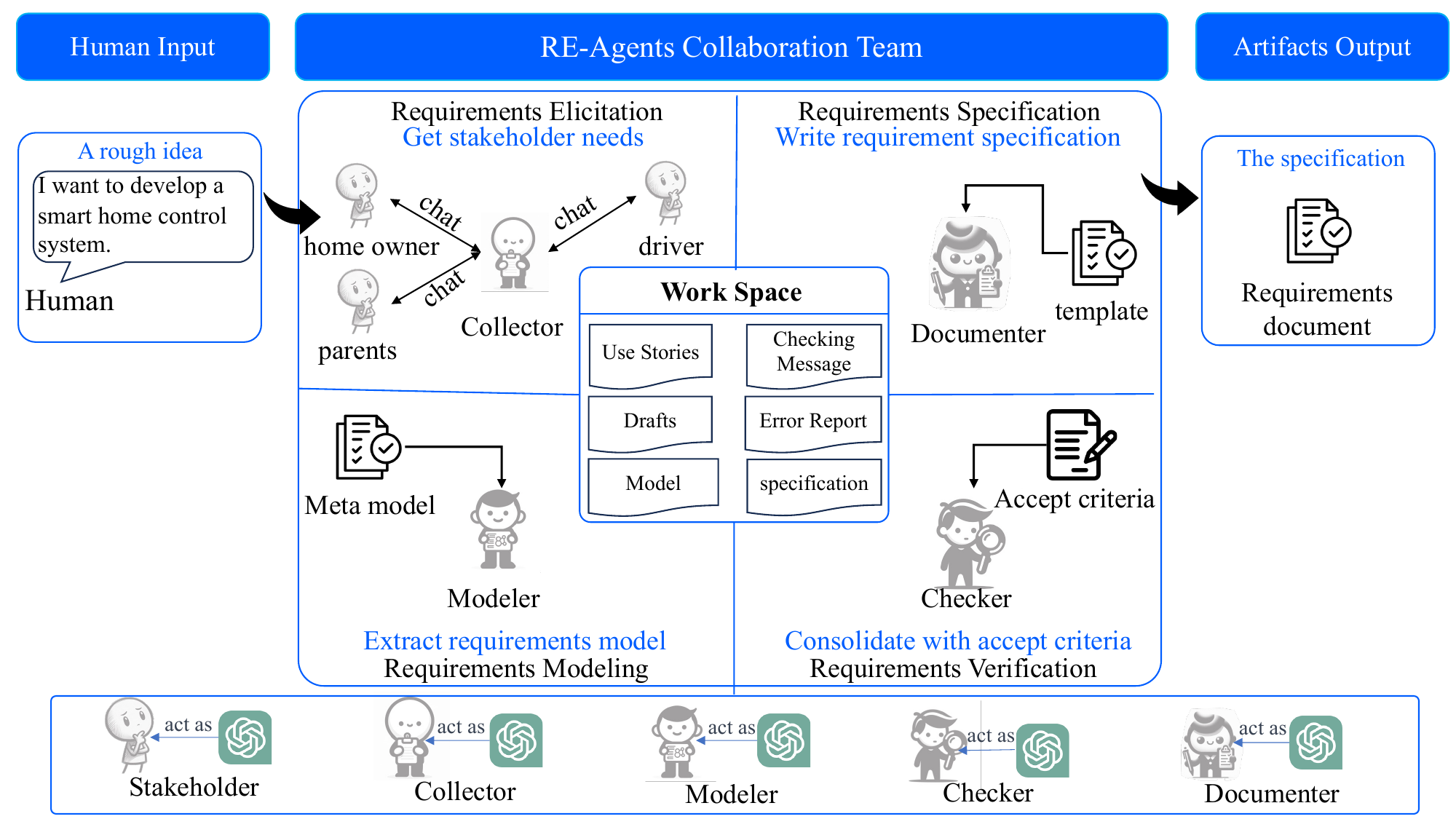}
    \caption{The overview of our MARE}
    \label{fig:enter-label}
\end{figure*}

Concretely, we design a multi-agent framework, \textit{MARE}, that leverages LLMs in the RE process. 
We design prompt-engineering-driven LLM agents to allow these agents with abilities to do RE tasks and introduce a shared workspace in which the artifacts can be stored as the communication structure. 
To obtain requirements specifications that meet quality standards, these agents work together iteratively and collaboratively to complete the RE process, just like the Human Requirements Engineering team does.
Figure \ref{fig:enter-label} depicts this framework, in which, we assume the following four tasks in the RE process:
%The four tasks work with collaboration 
%In this work, we decompose this system into four tasks, including requirement elicitation, modeling, verification and specification. The four tasks work with collaboration as show in Figure \ref{fig:enter-label}.
\begin{itemize}
    \item \textbf{Elicitation.} Given a rough idea of requirements $X$, this task collects stakeholders' needs $U$ and generates a requirements draft $D$. 
    \item \textbf{Modeling.} Based on the generated requirements draft $D$, this task generates a requirements model $M$ as required by the `Metamodel'.
    \item \textbf{verification.} The task detects requirements smells $S$ in the requirements draft $D$ and requirements model $M$ with `accept criteria'.
    \item \textbf{Specification.} If requirements smells $S$ don't exist, this task generates a requirement specification $R$ based on the requirements draft $D$, and requirements model $M$ by following a `Template'. Otherwise, this task generates an Error report $E$.
\end{itemize}

\subsection{Tasks} \label{module}
As shown in Figure \ref{fig:enter-label}, the RE-Agent Collaboration team conducts iteratively four tasks in \textit{MARE}.
%each task involves one or two specific agents with different roles. Here we introduce the design of agents involved in each task.

The purpose of the requirements elicitation task is to get stakeholders' needs $U$ corresponding to the rough idea of system requirements $X$. 
Various \textbf{Stakeholders} can be involved to express their requirements for this system.
The requirements \textbf{Collector} interviews these stakeholders to further delve into their needs and preferences and write down their needs into a requirement draft $D$. 
The information \textbf{Collector} agent and a set of \textbf{Stakeholders} agents are designed to complete this task.  

The purpose of the requirements modeling task is to elaborate the requirements models, including the extraction of the requirements entities and relations. 
Following the previous study\cite{JinWJ23}, the requirements \textbf{Modeler} identifies requirements entities from the requirements draft $D$, e.g., the \textit{actor}s in use case diagrams and the \textit{machine} in problem diagrams. 
The \textbf{Modeler} then decides whether a certain relationship exists between any pair of entities and determines the type of the relationship, e,g., the \textit{include} in use case diagrams and the \textit{requirements reference} in problem diagrams. 
Finally, the \textbf{Modeler} combines these entities and relationships to form the requirements model $M$. 
The \textbf{Modeler} agent is designed to accomplish this modeling task.

The purpose of the requirements verification task is to confirm that the system requirements contain all the necessary elements of well-written requirements, e.g., correctness, completeness, and consistency. 
In the real world, the quality \textbf{Checker} first figures out the acceptance criterion, then reads the requirements draft $D$ and requirements model $M$ to assess the quality of the current requirements draft. 
The \textbf{Checker} agent is designed in \textit{MARE} to finish this quality-checking task.

Finally, if the quality of the current version of the requirements document meet the quality criteria, the specification task write the software requirements specification (SRS),  otherwise presents an error report. 
The \textbf{Documenter} writes SRS $R$ based on requirements draft $D$ and requirements model $M$ according to the pre-defined SRS template or presents the Error Report. 
The \textbf{Documenter} agent is designed for this task.

When an error report appears in the shared workspace, \textbf{Collector} and \textbf{Modeler} will work sequentially, improving the requirements draft $D$ and requirements model $M$, respectively, and then \textbf{Checker} will check their quality again.

\subsection{Agent Definition} \label{agent}
The necessary agents in \textit{MARE} have been identified in Section \ref{module}. 
They are \textbf{Stakeholders}, \textbf{Collector}, \textbf{Modeler}, \textbf{Checker}, and \textbf{Documenter}. 
Here we describe each agent and the interactions among agents.

\begin{figure}
    \centering
    \includegraphics[scale=0.5]{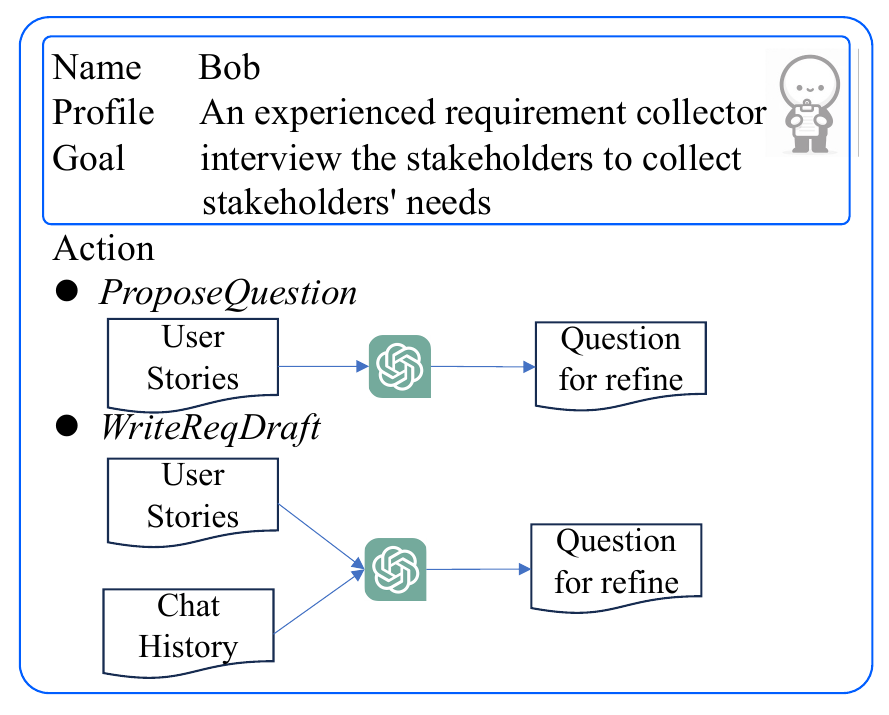}
    \caption{The profile of the \textbf{Collector} agent.}
    \label{agentprofile}
\end{figure}

\textbf{Agent Specialization.}
In \textit{MARE}, any agent has a role definition, which includes the name, the profile, the goal, and the actions. 
The name and profile indicate which role the agents play. 
The goal describes tasks for which the agent. 
The action gives the skills that the agent has. 
For instance, \textbf{Collector} can interview with \textbf{Stakeholders} and write the requirement draft based on user stories. 
Thus, the \textbf{Collector} agent is equipped with \textit{ProposeQuestion} action and \textit{WriteReqDraft} action. 
Figure \ref{agentprofile} shows an example of the \textbf{Collector} agent. 

The other agents are defined via the same way based on the analysis in section \ref{module}. 
The \textbf{Stakeholders} agent can express their expectations for the systems and answer questions from \textbf{Collector}. 
Then, So the \textbf{Stakeholders} agent can do actions such as \textit{SpeakUserStories} and \textit{AnswerQuestion}. 
The \textbf{Modeler} agent splits the requirements modeling task into two steps, which are extracting requirements entities and determining the relationships among the entities. 
So the \textbf{Modeler} agent can do actions such as \textit{ExtractEntity} and \textit{ExtractRelation}. 
The \textbf{Checker} agent can do \textit{CheckRequrement} action. 
The \textbf{Documenter} agent can write requirements specifications or present the error report. 
So \textbf{Documenter} agent is equipped with \textit{WriteSRS} and \textit{WriteChceckReport} action.
The role definition of each agent is shown in Table \ref{agentdefinition}.

 \begin{table*}[]
     \centering
     \caption{The role definition of agents}
     \begin{tabular}{lllll}
\hline
\textbf{Agent} & \textbf{Name} & \textbf{Profile}                                                                  & \textbf{Goal}                                                                                                 & \textbf{Action}                                                                       \\ \hline
Stakeholders   & Alice         & \begin{tabular}[c]{@{}l@{}}An experienced \\ requirement stakeholder\end{tabular} & \begin{tabular}[c]{@{}l@{}}express and task the stakeholders' need \\ for system to be developed\end{tabular} & \begin{tabular}[c]{@{}l@{}}a. \textit{SpeakUserStories}\\ b. \textit{AnswerQuestion}\end{tabular} \\ \hline
Collector      & Bob           & \begin{tabular}[c]{@{}l@{}}An experienced \\ requirement collector\end{tabular}   & \begin{tabular}[c]{@{}l@{}}interview the stakeholders to collect \\ stakeholders' needs\end{tabular}          & \begin{tabular}[c]{@{}l@{}}a. \textit{ProposeQuestion}\\ b. \textit{WriteReqDraft}\end{tabular}   \\ \hline
Modeler        & Carol         & \begin{tabular}[c]{@{}l@{}}An experienced \\ requirement modeler\end{tabular}     & \begin{tabular}[c]{@{}l@{}}extract requirement model, including  \\ entities and relations\end{tabular}       & \begin{tabular}[c]{@{}l@{}}a. \textit{ExtractEntity}\\ b. \textit{ExtractRelation}\end{tabular}     \\ \hline
Checker        & Dave          & \begin{tabular}[c]{@{}l@{}}An experienced \\ requirement checker\end{tabular}     & \begin{tabular}[c]{@{}l@{}}check the requirement quality based \\ the requirement model.\end{tabular}         & a. \textit{CheckRequirement}                                                                \\ \hline
Documenter     & Eve           & \begin{tabular}[c]{@{}l@{}}An experienced \\ requirement documenter\end{tabular}  & \begin{tabular}[c]{@{}l@{}}write the requirement specification or\\ or checking report.\end{tabular}         & \begin{tabular}[c]{@{}l@{}}a. \textit{WriteSRS}\\ b. \textit{WriteCheckReport}\end{tabular}      \\ \hline
\end{tabular}
     \label{agentdefinition}
 \end{table*}

\begin{figure}
    \centering
    \includegraphics[width=0.5\textwidth]{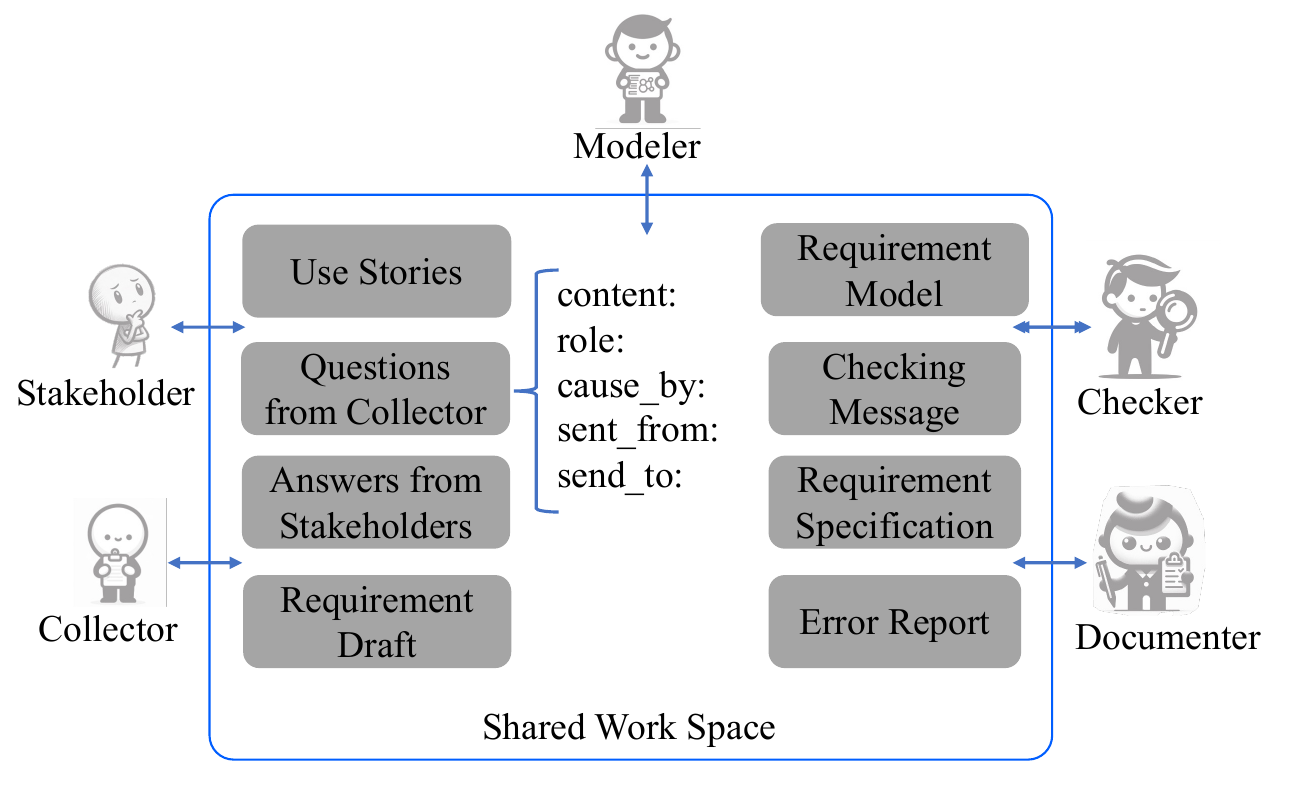}
    \caption{The mechanism of shared workspace.}
    \label{workspace}
\end{figure}

% 多专家系统
\textbf{Interaction between Agents.} \textit{MARE} adopts a shared workspace mechanism\cite{metagpt} to facilitate effective interaction and collaboration between different agents. 
As shown in figure \ref{workspace}, the shared workspace stores various requirements artifacts which can be operated by agents, such as the use stories, the requirements drafts, the requirements models, and the requirements specifications, etc.. 
The requirements artifacts in the workspace have five properties, which are content, role, caused\_by, sent\_from, and send\_to. 
The content property represents the content of the requirements artifacts. 
The role property indicates which agent generates the requirements artifacts. 
The caused\_by property describes which action generates the requirements artifacts, which can be used to perform action migration. 
The action migration mechanism will be described in detail in the next section. 
The sent\_from and send\_to properties give the flow of requirements artifacts among agents. 
For example, the content of use stories can be ``\textit{As a homeowner, I want to be able to control the temperature in my home remotely, so that I can adjust it to my desired level before I arrive.}'', 
the role and cause\_by of use stories can be stakeholders and \textit{SpeakUserStories}. 
The sent\_from and send\_to can be \textbf{Stakeholders} and \textbf{Collector}.

% The mechanism of the shared workspace allows the agents to access various requirements artifacts but also enables the agents to selectively receive requirements artifacts sent to themselves. 

\subsection{Action Definition} \label{actiondefinition}
\textbf{Action Space.} Section \ref{agent} describes the actions of each agent.
There are a total of nine actions in \textit{MARE}. 
Each action is implemented by designing a prompt for LLMs. 
Therefore, we designed multiple prompts of various actions for LLMs and the details about these prompts are at the open link\cite{material}. 
Table \ref{action_tables} shows the input and output for each action. 
The input will be put in each designed prompt and the output is referred to generated requirements artifacts by each action. 
The input for these actions is generated by agents within \textit{MARE}, except for the rough idea of requirements which is provided by humans. 
For the requirements draft and software requirement specification in Table \ref{action_tables}, their generation follows our predefined templates. 
The details of these templates are written in the prompt of \textit{WriteReqDraft} action and \textit{WriteSRS} action, which are publicly available at the same open link.

\begin{table}[]
    \centering
    \caption{The input and output artifacts for each action}
    \begin{tabular}{lll}
\hline
\textbf{Action}      & \textbf{Input}                                                                                                                          & \textbf{Ouput}                                                               \\ \hline
\textit{SpeakUserStories} & \begin{tabular}[c]{@{}l@{}}rough\_idea\_of\_requirements,\\ num\_user\_storeis\end{tabular}                                                              & user\_stories                                                               \\ \hline
\textit{ProposeQuestion}    & \begin{tabular}[c]{@{}l@{}}rough\_idea\_of\_requirements\\ user\_stories\end{tabular}                                                                    & requirement\_question                                                        \\ \hline
\textit{AnswerQuestion}     & \begin{tabular}[c]{@{}l@{}}rough\_idea\_of\_requirements,\\ requirement\_question\end{tabular}                                                           & stakeholders\_answer                                                         \\ \hline
\textit{WriteReqDraft}    & \begin{tabular}[c]{@{}l@{}}rough\_idea\_of\_requirements,\\ user\_stories,\\ requirement\_question,\\ requirement\_answer,\\ draft template\end{tabular} & requirement\_draft                                                           \\ \hline
\textit{ExtractEntity}     & \begin{tabular}[c]{@{}l@{}}requirement\_meta\_model,\\ requirement\_draft\end{tabular}                                                  & modeling\_entities                                                           \\ \hline
\textit{ExtractRelation}    & \begin{tabular}[c]{@{}l@{}}requirement\_meta\_model,\\ requirement\_draft,\\ modeling\_entities\end{tabular}                            & modeling\_relationship                                                       \\ \hline
\textit{CheckRequirement}   & \begin{tabular}[c]{@{}l@{}}modeling\_entities,\\ modeling\_relationship,\\ requirement\_draft\end{tabular}                              & checking\_message                                                            \\ \hline
\textit{WriteSRS}           & \begin{tabular}[c]{@{}l@{}}modeling\_entities,\\ modeling\_relationship,\\ requirement\_draft,\\ SRS\_template\end{tabular}             & \begin{tabular}[c]{@{}l@{}}software requirement\\ specification\end{tabular} \\ \hline
\textit{WriteCheckReport} & checking\_message                                                                                                                       & error\_report                                                                \\ \hline
\end{tabular}
    \label{action_tables}
\end{table}

\begin{figure}
    \centering
    \includegraphics[width=0.5\textwidth]{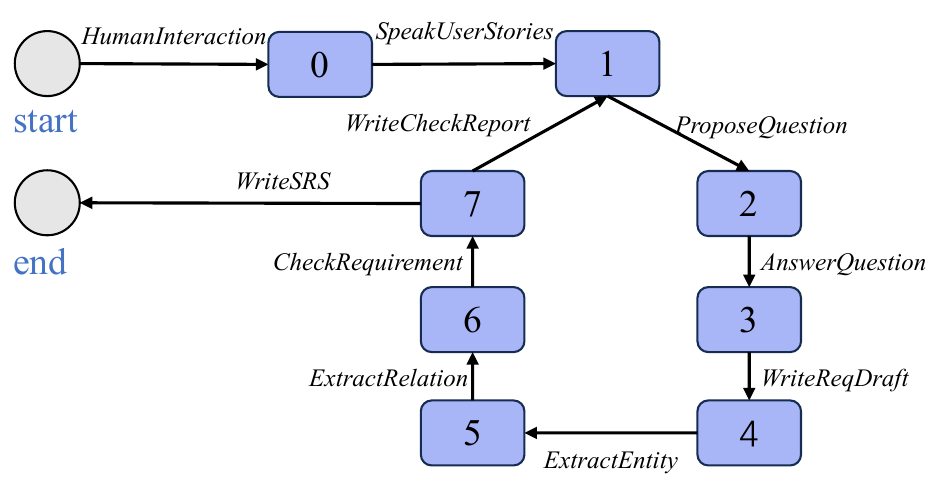}
    \caption{The action migration diagram in MARE}
    \label{action}
\end{figure}

\textbf{Action Migration.} \textit{MARE} iteratively performs a series of actions to achieve the end-to-end RE process. 
There is a sequential relationship between these actions. 
For example, action \textit{WriteDraft} happens after action \textit{SpeakUserStories}. 
Figure \ref{action} shows the whole action migration mechanisms in \textit{MARE}. 
\textit{HumanInteraction} is the start action and it can provide human feedback to MARE.

% 3页半
\section{Study Design}
To assess the effectiveness of our \textit{MARE}, we perform an extensive study to answer three research questions. 
In this section, we describe the details of our study, including evaluation cases, metrics, baselines, and experiment settings.

\subsection{Research Questions}
Our study aims to answer three research questions(RQ). 
In RQ1, we compare our \textit{MARE} to three SOTA automated requirements modeling baselines on 9 evaluation cases and 1 dataset to prove the modeling superiority of \textit{MARE}. 
In RQ2, we conduct a human evaluation to evaluate the generated software requirements specifications on 9 software systems in three aspects. 
In RQ3, we conduct an ablation study to prove the contributions of multi-agent collaboration.

\textbf{RQ1: How do the requirements models generated by \textit{MARE} compared to the SOTA approaches?} We first use \textit{MARE} to generate requirements models on 9 evaluation cases. 
Then, we use multiple metrics to evaluate the generated requirements models and compare them with the existing three SOTA baselines. 
We select multiple LLMs to verify that \textit{MARE} is effective for different LLMs.

\textbf{RQ2: How do the requirements specifications generated by \textit{MARE}?}. 
We first use \textit{MARE} to generate software requirements specifications on 9 evaluation cases. 
Then, we manually evaluate the generated requirements specifications in three aspects. 
Finally, we calculate the average results.

\textbf{RQ3: How does \textit{MARE} perform compared to individual LLM?} 
Our \textit{MARE} contains multiple LLMs-based agents. 
We assess the contributions of agents collaboration by comparing \textit{MARE} with individual LLMs on requirements modeling.

\subsection{Evaluation Cases} \label{evaluationcases}
We conduct experiments on 5 public evaluation cases for generating use case diagrams, 1 public dataset for generating goal models, and 4 new evaluation cases for generating problem diagrams collected by this work.

\textbf{Five public evaluation cases} are proposed for automated modeling use case diagrams. 
The 5 cases are the ATM system(ATM)\cite{vinay2009approach}, the cafeteria ordering system (COS)\cite{UmberBN11}, the library system (TLS)\cite{BajwaC11}, the assembly system(TAS)\cite{BajwaN11} and the time monitor system (TMS)\cite{UmberB12}. 
They are common and are often used to evaluate DL model to extract actors and actions in the use case. 
The original requirements descriptions for these cases can be obtained from the existing literature. 

\textbf{One public evaluation dataset} are proposed for automating modeling goal models\cite{Zhou0W22}. 
The origin requirements descriptions in this dataset come from PURE\cite{FerrariSG17}, a dataset of 79 publicly available natural language requirements documents collected from the Web. 
In this paper, we refer to this dataset as GoalModelDataset. 

\textbf{Four new evaluation cases} are proposed for automated modeling problem diagrams. 
The 4 cases are the smart home control system (TSHCS), the temperature control system (TTCS), the air conditioner control system (TACCS), and the humidistat control system (THCS). 
The origin requirements descriptions for these cases are shown in the same open link as the section \ref{actiondefinition}.

\subsection{Evaluation Metrics} \label{evaluationmetrics}
\textbf{Evaluation for requirements models.} We use three commonly used metrics to evaluate the performance, i.e., Precision, Recall, and F1. 
(1) Precision (P), which refers to the ratio of the number of correct predictions to the total number of predictions; 
(2) Recall (R), which refers to the ratio of the number of correct predictions to the total number of samples in the golden test set; and 
(3) F1, which is the harmonic mean of precision and recall. 
When comparing the performance, we care more about F1 since it is balanced for evaluation.

\textbf{Evaluation for requirements specifications.} 
Following the previous work\cite{Danning}, we manually evaluate the generated specifications by \textit{MARE} in three aspects, including completeness, correctness, and consistency. 
For each aspect, the score is ranging from 0 to 2 (from bad to good). 
The evaluators are computer science master students and are not co-authors.

\subsection{Baselines} \label{baselines}
We select 3 recently proposed automated requirements modeling approaches as baselines. 
They are state-of-the-art methods for automated modeling of problem diagrams, use case diagrams, and goal models, respectively.
\begin{itemize}
    \item \textbf{IT4RE}\cite{AlhroobIA18}: is an approach to automatically identify actors and actions from natural language requirements’ description using an NLP parser.
    \item \textbf{EPD}\cite{JinWJ23}: is a method for \textbf{E}xtracting requirements entities, e.g.,  machine domain, requirement domain and given domain, in \textbf{P}roblem \textbf{D}iagram based on BERT.
    \item \textbf{HAGM}\cite{Zhou0W22}: is a \textbf{H}ybrid \textbf{A}pproach to requirements entities, e.g., role, agent, and relation, in \textbf{G}oal \textbf{M}odel based on machine learning and logical reasoning. 
\end{itemize}
\begin{table*}[]
    \centering
    \caption{The results of MARE on modeling problem diagram}
    \begin{tabular}{ccccccccccccc}
\hline
                        & \multicolumn{3}{c}{EPD} & \multicolumn{3}{c}{MARE(gpt-3.5-turbo)}              & \multicolumn{3}{c}{MARE(text-davinci-002)} & \multicolumn{3}{c}{MARE(text-davinci-003)} \\ \cline{2-13} 
\multirow{-2}{*}{Cases} & P(\%)  & R(\%) & F1(\%) & P(\%)     & R(\%)      & F1(\%)                      & P(\%)        & R(\%)        & F1(\%)       & P(\%)        & R(\%)       & F1(\%)        \\ \hline
TSHCS                   & 76.4   & 67.9  & 72.7   & 85.7      & 92.3       & 83.8                        & 86.9         & 79.3         & 82.9         & 80.6         & 82.8        & 81.7          \\
TTCS                    & 72.5   & 58.3  & 64.6   & 81.6      & 86.1       & 83.8                        & 72.9         & 80.1         & 76.3         & 81.7         & 84.2        & 82.9          \\
TACCS                   & 71.3   & 62.4  & 66.6   & 81.3      & 88.4       &  84.7 & 86.6         & 85.7         & 86.1         & 85.3         & 82.5        & 83.9          \\
THCS                    & 74.7   & 67.2  & 70.8   & 81.8      & 86.5       & 84.1                        & 84.3         & 80.1         & 82.1         & 82.2         & 77.9        & 80            \\ \hline
Average                 & 73.7   & 63.9  & 68.7   & 82.6{\color[HTML]{FE0000}(8.9)} & 88.3{\color[HTML]{FE0000}(24.4)} & 84.1{\color[HTML]{FE0000}(15.4)}                  & 82.6{\color[HTML]{FE0000}(8.9)}    & 81.3{\color[HTML]{FE0000}(17.4)}   & 81.9{\color[HTML]{FE0000}(13.2)}   & 82.5{\color[HTML]{FE0000}(8.8)}    & 81.9{\color[HTML]{FE0000}(18)}    & 82.1{\color[HTML]{FE0000}(13.4)}    \\ \hline
\end{tabular}
    \label{problemdiagram}
\end{table*}

\begin{table*}[]
\caption{The Results of MARA on modeling Use case diagram}
\centering
\label{usecasediagram}
\begin{tabular}{ccccccccccccc}
\hline
                        & \multicolumn{3}{c}{IT4RE} & \multicolumn{3}{c}{MARE(gpt-3.5-turbo)}              & \multicolumn{3}{c}{MARE(text-davinci-002)} & \multicolumn{3}{c}{MARE(text-davinci-003)} \\ \cline{2-13} 
\multirow{-2}{*}{Cases} & P(\%)  & R(\%)  & F1(\%)  & P(\%)      & R(\%)     & F1(\%)                      & P(\%)         & R(\%)       & F1(\%)       & P(\%)         & R(\%)       & F1(\%)       \\ \hline
ATM                     & 50     & 83     & 63      & 72.9       & 83.3      & 77.8                        & 75.6          & 83.3        & 79.2         & 76.7          & 72.7        & 74.6         \\
COS                     & 63     & 56     & 59      & 79.2       & 80        & 79.6                        & 72.7          & 79.6        & 76           & 75.7          & 81.8        & 78.6         \\
TLS                     & 50     & 56     & 53      & 78.3       & 77.8      & 78                          & 74.8          & 74.8        & 74.8         & 71.4          & 78.9        & 75           \\
TAS                     & 17     & 100    & 29      & 74.9       & 82.8      & 78.7 & 72.3          & 78.7        & 75.3         & 77.8          & 81.1        & 79.4         \\
TMS                     & 56     & 100    & 71      & 81.6       & 79.1      & 80.3                        & 80.7          & 77.6        & 79.1         & 82.2          & 77.9        & 80           \\ \hline
Average                 & 47.2   & 79     & 55      & 77.4{\color[HTML]{FE0000}(30.2)} & 80.6{\color[HTML]{FE0000}(1.6)} & 78.9{\color[HTML]{FE0000}(23.9)}                  & 75.2{\color[HTML]{FE0000}(28.0)}    & 78.8{\color[HTML]{009901}(0.2)}   & 76.8{\color[HTML]{FE0000}(21.8)}   & 76.8{\color[HTML]{FE0000}(29.6)}    & 78.5{\color[HTML]{009901}(0.5)}   & 77.5{\color[HTML]{FE0000}(22.5)}   \\ \hline
\end{tabular}
\end{table*}

\begin{table*}[]
\caption{The results of MARE on Modeling Goal Model}
\centering
\label{goalmodel}
\begin{tabular}{cccccccccccccc}
\hline
\multirow{2}{*}{Datasets}         & \multirow{2}{*}{Extracted Elements} & \multicolumn{3}{c}{HAGM} & \multicolumn{3}{c}{MARE(gpt-3.5-turbo)} & \multicolumn{3}{c}{MARE(text-davinci-002)} & \multicolumn{3}{c}{MARE(text-davinci-003)} \\ \cline{3-14} 
                                  &                                     & P(\%)  & R(\%)  & F1(\%) & P(\%)       & R(\%)       & F1(\%)      & P(\%)        & R(\%)        & F1(\%)       & P(\%)        & R(\%)        & F1(\%)       \\ \hline
\multirow{2}{*}{GoalModelDataset} & Role                                & 80.8   & 83.9   & 82.3   & 81.3        & 84.3        & 82.7        & 81.1         & 83.4         & 82.3         & 82.3         & 83.5         & 82.9         \\
                                  & Action                              & 93.2   & 90.4   & 91.7   & 93.5        & 91.4        & 92.4        & 92.5         & 91.2         & 91.8         & 92.4         & 91.2         & 91.8         \\ \hline
\multicolumn{2}{c}{Average}                                             & 87     & 87.2   & 87     & 87.4        & 87.9        & 87.6        & 86.8         & 87.3         & 87.1         & 87.4         & 87.4         & 87.4         \\ \hline
\end{tabular}
\end{table*}
\subsection{Experiment Settings}
The implementation details of our \textit{MARE} are as follows. 
We use the open-source multi-agent framework - MetaGPT\cite{metagptwebsite} to build our \textit{MARE}. 
For the LLM engine of the \textit{MARE}, we use different versions of ChatGPT-3.5, named gpt-3.5-turbo \cite{models}, text-davinci-002 \cite{models} and text-davinci-003\cite{models}. 
They are the closed-source model developed by OpenAI. 
So we experimented by purchasing the access api-key from OpenAI. 
Considering that the responses of the LLMs are somewhat random, in order to ensure the stability of the experimental results, we set temperature as 0, max\_tokens as 3000, top\_p as 1, frequency\_penalty as 0, presence\_penalty as 0, best\_of as 1. 
These are the hyper-parameters of the LLMs. 

For requirements modeling, the experiments include generating modeling elements in problem diagrams, use case diagrams and goal models. 
To be specific, for the problem diagrams, we use \textit{MARE} to generate the the machine domains, the requirements domains, the physical devices, the sharing phenomenon, and the requirements references. 
For the use case diagram, \textit{MARE} is to generate the actors and the use cases. 
For the goal model, \textit{MARE} is used to generate the roles and the agents.

 % 4页半
\section{Results and Analyses}

In our first research question, we evaluate the performance of the requirements modeling by \textit{MARE} compared with the previous automated requirements modeling approaches.

\textbf{RQ1:How do the requirements models generated by \textit{MARE} compared to the SOTA approaches?}

\begin{table*}[]
\centering
\label{ablation}
\caption{The results of ablation study}
\begin{tabular}{cccccccccccccccc}
\hline
\multirow{2}{*}{Strategies} & \multicolumn{3}{c}{The ATM} & \multicolumn{3}{c}{The Cafeteria} & \multicolumn{3}{c}{The Library} & \multicolumn{3}{c}{The Asemantic} & \multicolumn{3}{c}{The Time Monitor} \\ \cline{2-16} 
                            & P       & R       & F1      & P         & R         & F1        & P         & R        & F1       & P         & R         & F1        & P          & R          & F1         \\ \hline
Individual LLM              & 73.1    & 81.4    & 77.0    & 79.7      & 77.6      & 78.6      & 78.9      & 75.2     & 77.0     & 75.1      & 81.4      & 78.1      & 79.2       & 77.2       & 78.2       \\ \hline
MARE                   & 72.9    & 83.3    & 77.8    & 79.2      & 80.0      & 79.6      & 78.3      & 77.8     & 78.0     & 74.9      & 82.8      & 78.7      & 81.6       & 79.1       & 80.3       \\ \hline
\end{tabular}
\end{table*}

\textbf{Setup.} We evaluate baselines (Section \ref{baselines}) and \textit{MARE} on 9 evaluation cases and 1 public evaluation dataset (Section \ref{evaluationcases}). 
The evaluation metrics are described in Section \ref{evaluationmetrics}, i.e., the Precision(P), Recall(R), and F1. 
For all metrics, higher scores represent better performance. 
Given the name of each evaluation case, \textit{MARE} can generate modeling elements. 
We compute these three metrics based on these modeling elements. 

\textbf{Results.} 
Table \ref{problemdiagram}, Table \ref{usecasediagram} and Table \ref{goalmodel} show the experimental results of the problem diagrams, the use case diagrams, and the goal models compared with three baselines. Red indicates improved, and green indicates decreased.

\textbf{Analyses.} 
\textit{MARE} achieves the best results among all baselines. 
For modeling problem diagrams, \textit{MARE}(gpt-3.5-turbo) improves the SOTA approach, i.e., EPD, by 8.9\% for average precision, by 24.4\% for average recall and by 15.4\% for average F1. 
% Besides, \textit{MARE}(text-davinci-002) improved the SOTA approach by 8.9\%, 17.4\%, 13.2\% for average precision, average recall and average F1. \textit{MARE}(text-davinci-003) improved the SOTA approach by 8.8\%, 18\%, 13.4\%. 
Similarly, for modeling use case diagrams, \textit{MARE}(gpt-3.5) outperforms the SOTA approach, i.e., IT4RE, by 30.1\% for average precision, 1.6\% for average recall, and 23.8\% for average F1. 
For modeling goal models, \textit{MARE}(gpt-3.5) improves the SOTA approach, i.e., HAGM, by 0.4\% for average precision, 0.7\% for average recall, and 0.6\% for average F1.
These results indicate \textit{MARE} can more accurately generate modeling elements in various requirements models.

% We believe that the performance advantage of MARE is mainly attributed to the 

\begin{center}
\fcolorbox{black}{gray!10}{\parbox{.9\linewidth}{\textbf{Answer to RQ1:} \textit{MARE} achieves the best results among all baselines. 
% In particular, 
The average F1 score of \textit{MARE} is 84.1\%, 78.9\% and 87.6\% on three requirement model, improving their SOTA approach by 15.4\%, 23.8\% and 0.6\%}}
\end{center}

In RQ2, we aim to evaluate the performance of requirement specification generated by \textit{MARE}.

\textbf{RQ2: How do the requirements specifications generated by \textit{MARE}?}

\textbf{Setup.} 
We evaluate \textit{MARE} on 9 evaluation cases (Section \ref{evaluationcases}). 
Given the name of these cases, we collect the requirements specifications generated by \textit{MARE}. 
To guarantee the correctness of the evaluation, we built an inspection team, which consisted of three master students. 
All of them are fluent English speakers and have done intensive research work with RE. 
Each student scored each of the 9 requirements specifications from three aspects (Section \ref{evaluationmetrics}). 
Finally, the average score was calculated.

\textbf{Results.} Table \ref{specification} shows the results of evaluation for requirements specifications generated by \textit{MARE} (gpt-3.5-turbo).

\begin{table}[]
    \centering
    \label{specification}
    \caption{The score of requirement specification}
    \begin{tabular}{cccc}
\hline
Evaluation Cases & completeness & correctness & consistency \\ \hline
TSHCS        & 0.87   &         1.53     &        1.54               \\
TTCS         & 0.89   &         1.62     &        1.66                 \\
TACCS        & 0.78    &        1.69      &       1.87                  \\
THCS         & 1.04   &         1.71     &        1.74                 \\
ATM          & 0.98   &         1.66     &        1.95                \\
COS          & 0.97   &         1.74     &        1.71               \\
TLS          & 1.02   &         1.81     &        1.82               \\
TAS          & 1.06   &         1.85     &        1.77               \\
TMS          & 1.21   &         1.77     &        1.87             \\ \hline
Average          &  0.98            &   1.92          & 1.98            \\ \hline
\end{tabular}
    \label{tab:my_label}
\end{table}

\textbf{Analyses.} 
\textit{MARE} can generate requirement specifications for various software systems. 
For the 9 various software systems, the completeness score of the generated specification is between 0.78 and 1.21. 
The correctness score of them is between 1.53 and 1.85. 
Similarly, the consistency score of them is between 1.54 and 1.95. 
For average score, \textit{MARE} achieves 0.98 for completeness, 1.92 for correctness, and 1.98 for consistency. 
These results imply that \textit{MARE} can effectively generate requirements specifications.
We also notice that the completeness aspect is slightly lower than the correctness and consistency. 
By investigating the causes, we believe that there are two reasons. 
First, the reason why evaluators give lower scores for the completeness is that they think each section of the generated requirement specifications is insufficient. 
Second, the length of text is reduced by LLMs when generating long text.

\begin{center}
\fcolorbox{black}{gray!10}{\parbox{.9\linewidth}{\textbf{Answer to RQ2:} \textit{MARE} can effectively generates requirement specification. To be specific, the average score of \textit{MARE} is 0.98, 1.92 and 1.98 on the three aspects.}}
\end{center}

\textbf{RQ3:How does MARE perform compared to individual LLM on requirements modeling?}

\textbf{Setup.}  
We compare \textit{MARE} with an individual LLM to generate requirements models on 5 public cases. 
For \textit{MARE}, we compute three metrics (Section \ref{evaluationmetrics}) based on extracted requirements modeling elements from the generated requirements drafts. 
For the individual LLM, we directly use one LLM to extract requirements modeling elements from the requirements drafts generated by \textit{MARE} and computer the same metrics. 
Besides, we use gpt-3.5-turbo as the individual LLM and select the use case diagrams.
 
\textbf{Results and Analyses.} 
Table \ref{ablation} shows the comparative results between \textit{MARE} and the individual LLM on 5 public cases. 
\textit{MARE} can generate more correct requirements modeling elements compared with the individual LLM on the 5 public cases. 
To be specific, \textit{MARE} outperforms the individual LLM by 0.2\% for average precision, 2\% for average recall and 1.1\% for average F1.

\begin{center}
\fcolorbox{black}{gray!10}{\parbox{.9\linewidth}{\textbf{Answer to RQ3:} \textit{MARE} achieves better results comparing with the individual LLM. 
This demonstrates the superiority of collaboration of multiple agents.}}
\end{center}
 % 5页半
% \input{chapters/6Discussion} %6页
\section{Related Works}

\textbf{Deep-Learning-based Requirements Engineering.} 
%Requirements Engineering (RE) is the process elicitation, modeling, validation and specification. 
With the rapid development of the deep learning technique, many researchers are devoted to applying DL to RE to improve the effectiveness of software development\cite{tdb}\cite{EzziniA0S22}\cite{AlhoshanFZ23}. 
% Existing works on DL-based RE focuses on the individual tasks in the RE life cycle separately\cite{tdb}\cite{EzziniA0S22}\cite{AlhoshanFZ23}.  
For \textit{Requirement elicitation} task, recent work has more focus on mining user needs\cite{WangWZMSW22} from the open source community\cite{gitter} and facilitating conversations\cite{DicklerDMWBC22}. 
For \textit{Requirement modeling} task, some researchers use pre-defined rules to extract modeling entities and relations\cite{ElbendakVR11}. Others focus on building neural network models\cite{LiYS0HPLP20}\cite{DevlinCLT19} to extract the requirements models. 
Recent works also have attempted to explore the ability of LLMs for requirements modeling\cite{RuanCJ23}.
For \textit{Requirement verification} task, \cite{MuSZZZ20} designed a tool to detect smell in requirements requests based on rules. Besides, many works focus on detecting ambiguous and inconsistent\cite{EzziniA0S22} requirements. 
Compared to these existing studies, our work leverages LLMs throughout the entire requirements engineering through collaboration, rather than just dealing with a single task in RE.

\textbf{LLM-Based Multi-Agent Frameworks.} 
LLMs have shown remarkable performance across a wide range of domains. 
\cite{yilundu} improves the reasoning accuracy by leveraging multi-agent debate. 
\cite{chatdev} proposed ChatDev, a virtual chat-powered software development company consisting of multiple agents based on LLMs. 
\cite{metagpt} introduce MetaGPT, an innovative meta-programming framework incorporating efficient human workflows into LLM-based multi-agent collaborations. 
Compared to these studies, our work focuses on tasks in RE. 
% Although ChatDev and MetaGPT also contain requirements tasks, it is limited to the requirements elicitation task.  %6页半
\section{conclusion}
%In this paper, we present MARE, a multi-agents collaboration framework, that can perform requirements elicitation, requirements modelling, requirements verification and requirements specification to further increase efficiency in handling requirements engineering tasks. 
In this paper, we propose MARE, a multi-agent collaboration framework based on LLMs, to improve the efficiency of processing requirements engineering tasks. 
In MARE, different agents perform requirements elicitation, requirements modeling, requirements verification, and requirements specification, respectively, and collaborate to generate the requirement specification.
When conducting these tasks, some external knowledge has been used, e.g. The Modeler extracts the requirements models according to the `Meta model'; the Checker checks the models based on `Accept criteria'; and the Documenter produces the specification in terms of the `Template'.
%MARE can directly generates various requirement model (such as use case diagram and problem diagram) and solve the problem of proposing a method for each requirement model. 
MARE can contribute to multiple LLMs (such as gpt-3.5-turbo). 
We demonstrate the superiority of MARE's requirements modeling capabilities by comparing three SOTA automated requirements modeling approaches. 
Besides, we have provided insights into the quality of the requirements specification generated by MARE.
 % 7页
\bibliographystyle{IEEEtran}
\bibliography{IEEEabrv,myrefs}

\end{document}